\documentclass[a4paper,11pt]{article}
\usepackage{pos}

\title{Heavy flavor transport in the QGP medium}

\author*[a]{Bruno Scheihing-Hitschfeld}

\affiliation[a]{Kavli Institute for Theoretical Physics, University of California, Santa Barbara, \\
  California 93106, USA}

\emailAdd{bscheihi@kitp.ucsb.edu}

\abstract{I review recent developments on heavy flavor transport in the QGP medium, along two directions. The first is the transport of individual open heavy quarks. Leveraging the tools of heavy quark effective theory, recent work revealed a novel connection between the evolution equation of the heavy quark phase space distribution, the conditions for kinetic equilibration to take place, and a non-perturbatively defined expectation value of a Wilson loop characterized by the heavy quark velocity. Using these developments, I discuss results from a first exploration of the ensuing stopping and equilibration dynamics of heavy quarks in a strongly coupled environment. The second direction is the transport of quarkonia, bound states of heavy quarks, where novel Generalized Gluon Distributions have been recently defined and calculated in the framework of potential non-relativistic QCD. I discuss their application to the dynamics of quarkonium in heavy-ion collisions, highlighting the need to account for non-Markovian effects in the interactions between the heavy quark pair and quark-gluon plasma --- which have thus far resisted a systematic characterization.}

\FullConference{The Thirteenth Annual Large Hadron Collider Physics (LHCP2025)\\
5-9 May 2025\\
Taipei, Taiwan\\}

\begin{document}
\maketitle

Heavy quarks and quarkonia are amongst the most informative probes of quark-gluon plasma (QGP). Because heavy quarks are long-lived relative to the time interval during which QGP exists in a heavy-ion collision, they can be identified experimentally and their corresponding hadronic yields can be compared with those in colliding systems in which no QGP is formed, allowing one to obtain direct evidence of medium-induced modifications. 
As such, the dynamics of heavy quarks provides a unique window into the strongly coupled physics of QGP. 

The fact that medium modifications have been observed and precisely measured for individual heavy quarks (e.g.,~\cite{CMS:2017qjw,PHENIX:2022wim,ALICE:2023gjj}) and for quarkonium (e.g.,~\cite{CMS:2023lfu,ALICE:2023hou,LHCb:2024hrk,STAR:2025imj}) in heavy-ion collisions presents a unique opportunity to learn about the detailed structure and dynamics of hot QCD matter.
In what follows, I discuss selected aspects of the theoretical description of these phenomena, with emphasis on recent theoretical developments that aim to connect experimental data directly with quantities that have been formulated (and sometimes calculated) in QCD, thus allowing us to directly test our understanding of the dynamics of the strong force under extreme conditions.

\section{Open Heavy Flavor in QGP}

The main quantity of interest to characterize transport of individual heavy quarks in medium is the heavy quark (HQ) phase space distribution $\mathcal{P}({\bf x},{\bf p},t)$, which describes how many such particles can be found at position ${\bf x}$ with momentum ${\bf p}$ at time $t$. 
Medium effects on heavy quark propagation, and therefore properties of QGP, can be inferred by comparing the evolution of $\mathcal{P}$ at nonzero temperature $T>0$ with its evolution in vacuum. In what follows, I will discuss how this takes place in a fully thermalized, spatially homogeneous QGP background. 



A simple model is to consider heavy quarks evolving according to Langevin dynamics with a Gaussian random force, which can be cast in terms of a Fokker-Planck equation
\begin{equation}
    \partial_t \mathcal{P} = \partial_i \! \left( \eta_D(p) p_i \mathcal{P} \right) + \frac12  \partial_i \partial_j \! \left( \left[ \kappa_T(p) \delta_{ij} + \left(\kappa_L(p) - \kappa_T(p) \right) \hat{p}_i \hat{p}_j \right]  \mathcal{P} \right) 
\end{equation}
which is characterized by three momentum-dependent transport coefficients: the drag coefficient $\eta_D(p)$ and the momentum broadening coefficients $\kappa_L(p)$, $\kappa_T(p)$ describing longitudinal and transverse momentum broadening, respectively. At $p = 0$, they are all related
\begin{equation}
    \kappa_T(0) = \kappa_L(0) = 2 T M \eta_D(0) \, ,
\end{equation}
where the first equality is simply a consequence of isotropy of the thermal background, and the second is a consequence of the fluctuation-dissipation theorem. Taking this model as a baseline, one can constrain these coefficients by comparing with data (see~\cite{Rapp:2018qla} for a review).

In this class of models, in order to guarantee the eventual kinetic equilibration of heavy quarks, the Einstein relation $\kappa_L(p) = 2 T E(p) \eta_D(p)$ needs to be imposed\footnote{This guarantees that the equilibrium distribution is $\propto \exp(- E(p)/T)$, up to terms suppressed in the heavy quark limit by $T/M$.}. While appealing because of its simplicity, this is actually a drawback of this approach if one wants to draw conclusions about QCD: in the heavy quark limit ($M \to \infty$) all of these transport coefficients can be formulated and in principle calculated directly in the quantum field theory that describes the thermal environment, and these have been shown to \textit{not} satisfy the Einstein relation~\cite{Moore:2004tg,Gubser:2006nz}.

Another approach is to study heavy quarks in a Boltzmann transport model $p^\mu \partial_\mu \mathcal{P} = C[\mathcal{P}]$, where the information about HQ-QGP interactions is encoded in the collision kernel
\begin{equation}
    C[\mathcal{P}]({\bf p} ) = \int d^3k \left[ \gamma({\bf p}+{\bf k},{\bf k} ) \mathcal{P}({\bf p}+{\bf k}) - \gamma({\bf p},{\bf k}) \mathcal{P}({\bf p}) \right] \, ,
\end{equation}
where $\gamma({\bf p},{\bf k})$ is the rate of collisions changing the HQ momentum from ${\bf p}$ to ${\bf p} - {\bf k}$. This rate is much more versatile (but also more complicated) than the three transport coefficients of Fokker-Planck dynamics, and --- once an input functional form of $\gamma({\bf p},{\bf k})$ is given --- it allows one to benchmark the applicability of Fokker-Planck descriptions. For more discussion on this, see~\cite{Das:2013kea}.

\begin{figure}[t]
    \centering
    \includegraphics[width=0.95\linewidth]{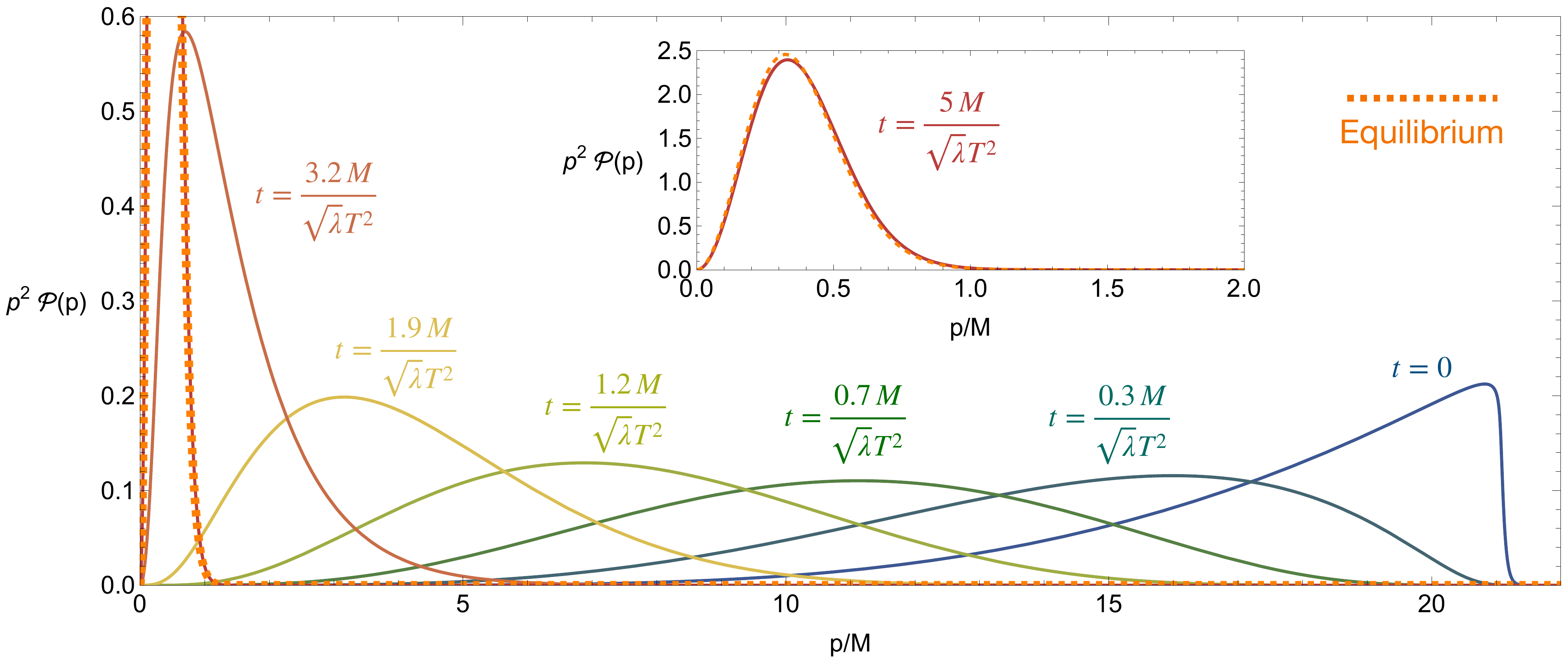}
    \caption{Time evolution of the heavy quark phase space distribution $\mathscr{P}$ for a selected initial condition, peaked around $p = 20 M$. The ratio of the mass over the temperature was chosen to be $M/T = 20$ (typical for a $b$ quark in a heavy ion collision). Different colors correspond to different snapshots of the time evolution.}
    \label{fig:example}
\end{figure}

Recently, we proposed to use Heavy Quark Effective Theory (HQET) to directly calculate the momentum transfer probability distribution from a HQ with velocity ${\bf v}$ to the medium in which it propagates~\cite{Rajagopal:2025rxr}. The answer can be written in terms of a Wilson loop $\langle W_{\bf v} \rangle ({\bf L})$
\begin{align}
    P({\bf k};{\bf v}) &= \frac{1}{(2\pi)^3} \int d^3{\bf L} \, e^{-i {\bf k} \cdot {\bf L}  } \langle W_{\bf v} \rangle ({\bf L}) \, , \label{eq:P-from-W}
\end{align}
made up by two long antiparallel Wilson lines (one coming from the amplitude for the HQ to change its momentum by $-{\bf k}$ and the other from the complex conjugate amplitude) separated by a distance ${\bf L}$. This probability distribution may be used to derive a \textit{Kolmogorov} equation
\begin{equation}
    \partial_t \mathcal{P} = - T \, K({\partial}_{\bf p} , {\bf p} ) \mathcal{P} \, , \label{eq:Kolmogorov}
\end{equation}
where the kernel $K$ is explicitly determined by the Wilson loop. Explicitly,
\begin{equation}
    K({\bf y}, {\bf p}) = -\frac{1}{tT} \log \left[ \langle W[\mathcal{C}] \rangle_T \left({\bf L} = - i {\bf y}; {\bf v} = {\bf v}({\bf p}) \right) \right] \, ,
\end{equation}
where ${\bf v}({\bf p})$ is the heavy quark velocity for a corresponding momentum ${\bf p}$. Crucially, due to a KMS relation that the Wilson loop satisfies $\langle W_{\bf v} \rangle ({\bf L}) = \langle W_{\bf v} \rangle (-{\bf L} + i {\bf v}/T )$, $K$ is guaranteed to drive the HQ distribution $\mathcal{P}$ to kinetic equilibrium~\cite{Rajagopal:2025rxr}. 

Therefore, in this formulation, characterizing HQ transport is equivalent to characterizing $K$. In principle, one would calculate this quantity in QCD and use it to make predictions. However, the scarcity of non-perturbative methods to carry out QCD calculations in real time has made this task not yet feasible. Nonetheless, it is possible to inform our expectations by looking at similar theories, such as $\mathcal{N}=4$ SYM, where the Wilson loop $\langle W_{\bf v} \rangle ({\bf L})$ has been calculated~\cite{Rajagopal:2025ukd}. This has allowed us to calculate how the heavy quark phase space distribution would evolve in a strongly coupled $\mathcal{N}=4$ thermal plasma. See Figure~\ref{fig:example} for an example. A systematic investigation of the phenomenology for realistic initial conditions (i.e., the perturbatively produced spectrum of heavy quarks at the initial time in a heavy-ion collision) using the results from this theory is underway.




\section{Quarkonium in QGP}

Quarkonium is a multi-scale probe of QGP. In addition to their respective heavy quark masses, bottomonium ($b\bar{b}$) and charmonium ($c\bar{c}$) states are  characterized by their sizes and binding energies. This makes the physical process of interest qualitatively different to the propagation of a single, open heavy quark. Because both bottomonium and charmonium can exist in a variety of different states (analogously to a hydrogen atom), quarkonium states have nontrivial dynamics of their own even in the absence of a thermal environment. It is thus the modification of the quantum dynamics of quarkonium which informs our knowledge of QGP. In the non-relativistic $v \ll 1$ and small bound state size $r T \ll 1$ limits, the effective field theory that describes this dynamics is potential non-relativistic QCD (pNRQCD)~\cite{Brambilla:1999xf}.

To study this phenomenon, the appropriate theoretical approach is the Open Quantum Systems (OQS) framework. This formalism makes it possible to couple pNRQCD to the light degrees of freedom of QCD at finite temperature, and therefore study the medium modifications of quarkonium dynamics. Concretely, the dynamics of quarkonium is encoded in the reduced density matrix
\begin{align}
    \rho_{Q\bar{Q}}(t) = {\rm Tr}_{\rm QGP} \left[ U(t) \rho_{\rm tot}(t=0) U^\dagger(t) \right] \, . \label{eq:oqs-general}
\end{align}
The time evolution operator $U$ describes the dynamics of the whole system: the QGP environment, a heavy quark--antiquark pair, and the interaction between QGP and the $Q\bar{Q}$ pair. pNRQCD describes the latter two pieces of the dynamics, and the light QCD degrees of freedom describe the former.
Likewise, $\rho_{\rm tot}(t=0)$ is the initial density matrix of the whole system. For a review, see~\cite{Yao:2021lus}.

This framework allows one to formulate the gauge-invariant correlation functions --- which have been given the name of \textit{Generalized Gluon Distributions} (GGDs)~\cite{Nijs:2023dbc}  --- that one needs to calculate to describe in-medium quarkonium dynamics, or conversely, that can be extracted or constrained from quarkonium suppression data. These distributions have been studied analytically in QCD up to NLO~\cite{Binder:2021otw,Scheihing-Hitschfeld:2022xqx} and in $\mathcal{N}=4$ SYM in the strong coupling limit~\cite{Nijs:2023dks}.
Their definition in terms of chromoelectric field correlators is given by
\begin{align} 
\label{eq:g++-definition}
[g_{\rm adj}^{++}]^>(t) &\equiv \frac{g^2 T_F }{3 N_c}  \big\langle E_i^a(t)W^{ac}(t,+\infty) 
W^{cb}(+\infty,0) E_i^b(0) \big\rangle_T \, , \\ \label{eq:g---definition}
[g_{\rm adj}^{--}]^>(t) &\equiv \frac{g^2 T_F }{3 N_c} \big\langle W^{dc}(-i\beta - \infty, -\infty)
W^{cb}(-\infty,t)  E_i^b(t)
E_i^a(0)W^{ad}(0,-\infty)  \big\rangle_T  \, .
\end{align}
They have also been formulated in terms of correlation functions that are calculable in Euclidean QCD~\cite{Scheihing-Hitschfeld:2023tuz}. More recently, a lattice study of these correlators in Euclidean time was carried out~\cite{Brambilla:2025cqy}. 

These studies in Euclidean QCD are crucial as they provide invaluable non-perturbative information on the dynamics of quarkonium. Ideally, one would like to reconstruct their real-time counterparts in a frequency range around $\omega = 0$ that extends at least up to $|\omega| \sim 1$ GeV, so as to include all of the binding energies of $b\bar{b}$ and $c\bar{c}$ states. 
While the direct application of these Euclidean correlation functions to quarkonium dynamics in medium still requires the development of additional theoretical tools -- mostly due to the difficulties posed by the need to analytically continue them into real time -- the results in~\cite{Brambilla:2025cqy} are already indicative of the relevance of physical processes that have yet to be accounted for in current phenomenological descriptions. This is apparent in the transport equations where the GGDs appear, which are schematically of the form
\begin{equation}
    \frac{d\rho_{mn}}{dt} = -i (H_{mk} \rho_{kn} - \rho_{mk} H_{kn} ) + \int_0^t dt' M_{mnij} [g_{\rm adj}](t,t') \rho_{ij}(t') \, ,
\end{equation}
because in one way or another (see, e.g., the quantum Brownian motion limit~\cite{Brambilla:2016wgg-2017zei} or the quantum optical limit~\cite{Yao:2018nmy}), practical applications of this formula have so far relied on simplifying its right hand side by taking a Markovian limit, i.e., making this equation local in time (where $\rho$ only appears as $\rho(t)$ at a unique time $t$). However, the Markovian limit misses a wealth of information encoded in the GGDs: because it is a local approximation, it contains no information about possible asymmetries under time reversal in the GGDs. This is a problem, because the GGDs are in fact not invariant under time reversal: perturbative~\cite{Binder:2021otw}, strongly coupled~\cite{Nijs:2023dks}, and now lattice calculations~\cite{Brambilla:2025cqy} verify this. Furthermore, it has been shown in strongly coupled setups that, in fact, \textit{all} of the quarkonium--QGP interactions may be encoded in non-Markovian effects~\cite{Nijs:2023dbc}.

A systematic study of the non-Markovian dynamics of quarkonium in QGP using the GGDs calculated in QCD or in $\mathcal{N}=4$ has yet to be carried out. In light of the above discussion, such a study will likely allow one to draw novel conclusions about QCD at finite temperature from quarkonium suppression data in heavy-ion collisions.

\section{Outlook}

I have given an overview of recent theoretical developments on heavy flavor transport using effective field theories of QCD. These developments have brought novel, non-perturbatively defined field-theoretical quantities to the forefront of the study of heavy flavor propagation, unveiling new ways to directly connect the QCD Lagrangian with empirical observations. In tandem with the effective field theory framework, these quantities --- namely, the kernel $K$ of the Kolmogorov equation~\eqref{eq:Kolmogorov} and the GGDs~\eqref{eq:g++-definition} \&~\eqref{eq:g---definition} --- provide a testable, systematically improvable approach to characterize the interaction of heavy quarks with QGP and the properties thereof.

Beyond their application to describe the phenomenology of heavy-ion collisions, the fact that these developments provide an explicit, interpretable connection with field-theoretic quantities is sure to deepen our understanding of hot QCD matter and, more generally, non-equilibrium phenomena in quantum field theory.



\vspace{0.4cm}

{\small The work of BSH is supported by grant NSF PHY-2309135 to the Kavli Institute for Theoretical Physics (KITP), and by grant 994312 from the Simons Foundation.}

\end{document}